\documentclass[a4paper]{article}

\usepackage{ISCSLP2022}
\usepackage{graphicx}
\usepackage{adjustbox}
\usepackage{multirow}
\usepackage{bm}
\usepackage{enumitem}
\usepackage{url}
\usepackage{footmisc}
\usepackage[T1]{fontenc}
\title{APCodec+: A Spectrum-Coding-Based High-Fidelity and High-Compression-Rate Neural Audio Codec with Staged Training Paradigm}
\name{Hui-Peng Du, Yang Ai, Rui-Chen Zheng, Zhen-Hua Ling}
\address{National Engineering Research Center of Speech and Language Information Processing, \\University of Science and Technology of China, Hefei, P. R. China }
\email{redmist@mail.ustc.edu.cn, yangai@ustc.edu.cn, zhengruichen@mail.ustc.edu.cn, zhling@ustc.edu.cn}

\begin{document}
\maketitle
\begin{abstract}
This paper proposes a novel neural audio codec, named APCodec+, which is an improved version of APCodec. 
The APCodec+ takes the audio amplitude and phase spectra as the coding object, and employs an adversarial training strategy. 
Innovatively, we propose a two-stage joint-individual training paradigm for APCodec+. 
In the joint training stage, the encoder, quantizer, decoder and discriminator are jointly trained with complete spectral loss, quantization loss, and adversarial loss. 
In the individual training stage, the encoder and quantizer fix their parameters and provide high-quality training data for the decoder and discriminator. 
The decoder and discriminator are individually trained from scratch without the quantization loss. 
The purpose of introducing individual training is to reduce the learning difficulty of the decoder, thereby further improving the fidelity of the decoded audio. 
Experimental results confirm that our proposed APCodec+ at low bitrates achieves comparable performance with baseline codecs at higher bitrates, thanks to the proposed staged training paradigm. 

%We utilize a staged training strategy, starting with jointly training the encoder, quantizer, decoder, and discriminator in the first stage. Subsequently, we frozen the parameters of the encoder and quantizer while randomly initializing the parameters of the decoder and discriminator. The second stage concentrates on training the parameters of the decoder and discriminator exclusively. Intra-model experiment demonstrates the effectiveness of the training strategy proposed by APCodec+, as it outperforms APCOdec at 6 kbps when operating at 4.5 kbps. Inter-model experiment shows that APCodec+ is capable of maintaining high-quality speech generation at the low bitrate of 4.5 kbps, surpassing the performance of most baseline models at 12 kbps.
%Experimental results show that our proposed BiVocoder can achieve high-quality speech waveform generation in the analysis-synthesis task. 
%Furthermore, the features extracted by BiVocoder are demonstrated suitable for direct prediction by acoustic models. 

% Experimental results show that our proposed BiVocoder achieves comparable performance in speech quality and faster inference speed to the mainstream two-stage TTS models with mel-scale features on both 
\end{abstract}
\vspace{-1mm}
\section{Introduction}
\vspace{-1mm}
Neural audio codec utilizes neural networks to encode input audio into discrete representations, which can be reconstructed into a high-fidelity audio waveform by a decoder. 
This technology is commonly employed in the fields of communication and audiovisual compression \cite{brandenburg1994iso,tremain1976linear,kroon1986regular,salami1994toll}. There have also been some efforts leveraging audio codecs combined with large language models to achieve zero-shot text-to-speech (TTS) \cite{borsos2023audiolm,wang2023neural,zhang2023speechtokenizer,huang2023repcodec,ren2024fewer}.
The quality of the decoded audio and the bitrate of the discrete representation are two main metrics for evaluating the audio codec. 
The quality of decoded audio reflects the ability of the audio codec to reconstruct the original audio, while the bitrate reflects the compression capability of the audio codec. 
An ideal audio codec should synthesize high-quality and high-fidelity audio with a high compression rate (i.e., low bitrate).

Typically, a neural audio codec consists of an encoder, a quantizer, and a decoder. 
The encoder transforms input audio or features into intermediate representations, while the quantizer utilizes quantization strategies such as residual vector quantization (RVQ) \cite{vasuki2006review}, group-residual RVQ (GRVQ) \cite{yang2023hifi} and finite scalar quantization (FSQ) \cite{mentzer2023finite} to represent these intermediates in a discrete format, aiming to minimize the gap between the original and quantized features. 
Subsequently, the decoder reconstructs time-domain waveforms from the quantized features.

In general, audio codecs can be categorized into two main types: parametric codecs and waveform codecs. 
Parametric codecs utilize frame-level features extracted from audio such as linear predictive coding (LPC) \cite{o1988linear}, EVS \cite{dietz2015overview} and Opus \cite{valin2013high}. 
These representations, with lower frame rates compared to the original time-domain signal, are more suitable for compression by audio codecs. 
However, the audio quality produced by these codecs is relatively low and susceptible to noise interference. 
Waveform codecs directly manipulate the input audio and recent works like SoundeStream \cite{zeghidour2021soundstream}, Encodec \cite{defossez2023high}, and HiFi-Codec \cite{yang2023hifi}, employ this coding type, which does not require manually defined features. 
This learning-based approach often yields higher-quality decoded audio. 
However, direct manipulation at the waveform level, as opposed to frame-level processing, places certain demands on device computational power and challenges for high compression rate.

 In order to maximize the advantages of both parametric codecs and waveform codecs, our previous work, APCodec \cite{ai2024apcodec} utilized the amplitude and phase spectra obtained through short-time Fourier transform (STFT) of waveforms as coding object. After quantization by the quantizer, the decoder then restored the amplitude and phase spectra, which were subsequently transformed back into the time-domain waveform using inverse STFT (iSTFT). This all-frame-level operation significantly facilitates a reduction in the bitrate of discrete representations and enhances computational speed. 

Most neural audio codecs adopt a general training approach, i.e., jointly training the encoder, quantizer and decoder. 
However, this approach is unfriendly for the training of the decoder. 
This is because, in the early stages of training, the coarse encoder and quantizer provide poor-quality training data for the decoder, making the decoder's training process more challenging. 
%However, this approach may increase the learning cost of the decoder, as the outputs of the encoder and quantizer may differ while training.  
AudioDec \cite{wu2023audiodec} proposed a two-stage training strategy to address this issue. 
%AudioDec made changes to its training paradigm, dividing the training into two stages. 
For the best system mentioned in \cite {wu2023audiodec}, in the first stage, only the metrics loss is used to train the encoder, quantizer, and decoder, without training the discriminator. 
In the second stage, the encoder and quantizer are fixed, and a HiFi-GAN vocoder \cite{kong2020hifi} as well as the discriminator are trained. 
However, in the first stage of AudioDec, the absence of adversarial training loss affects the quality of training data for the HiFi-GAN vocoder in the second stage. Additionally, the introduction of an extra vocoder increases the model complexity. 
%However, AudioDec requires a vocoder to reconstruct high-quality audio.

\begin{figure*}[t]
	\centering
	\includegraphics[width=0.9\linewidth]{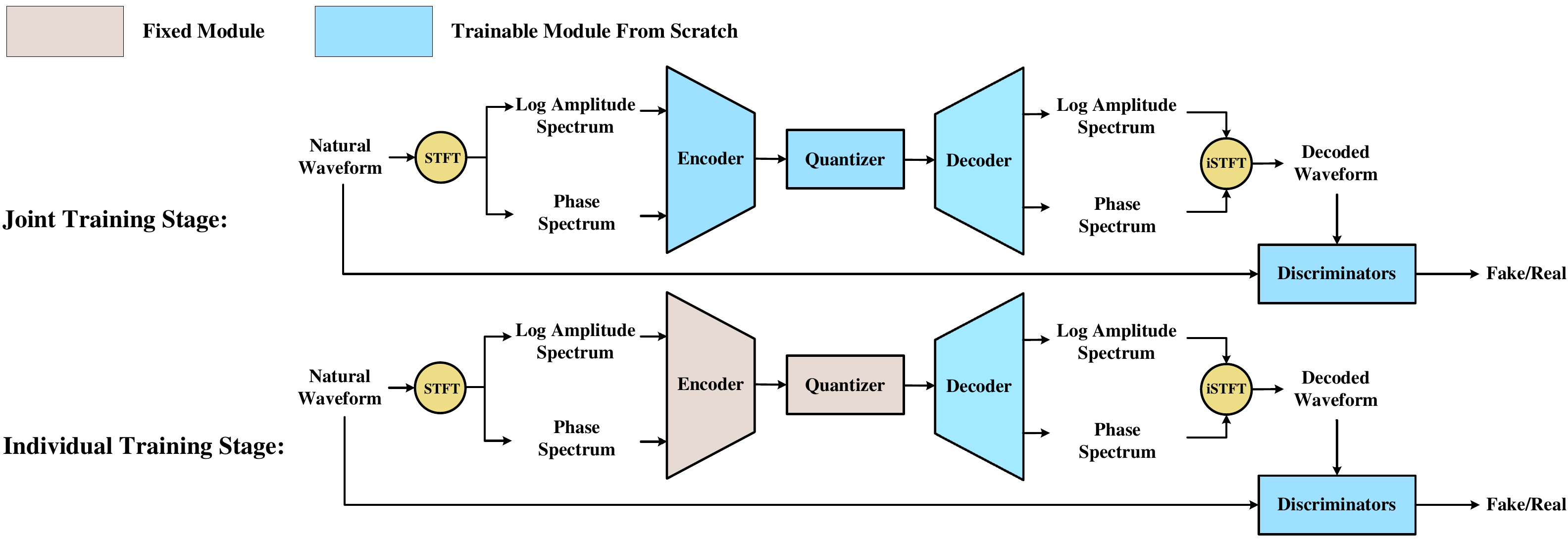}
	\caption{Illustration of the staged training paradigm for APCodec+. %Parameters for trainable modules would be randomly initialized between the two stages.
} \label{fig2}
\end{figure*}

Due to the impressive performance of APCodec \cite{ai2024apcodec}, we choose it as the backbone and propose an improved version, named APCodec+. 
The APCodec+ aims to alleviate the learning difficulties of the decoder and enhance the decoder's expressive capacity to better restore acoustic details regarding amplitude and phase spectra, thereby improving the quality of decoded audio. 
Specifically, the APCodec+ adopts a staged training paradigm, which is divided into joint and individual training stages. 
%We proposed APCodec+ aiming to enhance the decoder's expressive capacity to better restore acoustic details regarding amplitude and phase spectra, thereby improving the quality of synthesized speech. Therefore, we adopted a staged training approach. 
In the joint stage, the APCodec+ jointly trains the encoder, quantizer, decoder and discriminator. 
This allows the encoder and quantizer to produce sufficiently high-quality and content-rich quantized features, unlike AudioDec \cite{wu2023audiodec}, which relies solely on metric loss (i.e., without training the discriminator).
%. This allowed the encoder to learn feature extraction from the waveform level, rather than solely relying on metrics loss as in AudioDec. 
In the individual stage, the quantized features extracted by the fixed-parameter encoder and quantizer are used as the input of the decoder. 
The decoder is then reinitialized and individually trained from scratch. 
Unlike AudioDec \cite{wu2023audiodec}, APCodec+ does not require an additional vocoder for assistance. 
Experimental results confirm that at a 48 kHz sampling rate, APCodec+ achieves better decoded audio quality at 4.5 kbps than the original APCodec at 6 kbps. 
This indicates that the introduction of the staged training paradigm saves at least 1.5 kbps in bitrate (equivalent to approximately a 1.3$\times$ increase in compression rate). 
Compared to other well-known baseline audio codecs, such as SoundStream, Encodec, HiFi-Codec and AudioDec,  our proposed APCodec+ at 4.5 kbps is even capable of producing audio quality comparable to baselines at 12 kbps.

% we discarded the parameters learned by the decoder and discriminator in the first stage, reinitialized them, and used the quantized features extracted by the fixed-parameter encoder and quantizer as input of decoder. We then trained the decoder and discriminator from scratch to enhance the decoder's ability to restore acoustic details.

% The experiment demonstrates that our strategy, when applied to audio at 48 kHz and under 6 kbps conditions, shows improvements in amplitude spectrum, phase spectrum, and waveform compared to APCodec. Additionally, the quality of waveform generation can be maintained even when the bitrate is reduced to 4.5kbps. Compared to other well-known speech codecs, APCodec+ is capable of producing speech quality comparable to other codecs at 12 kbps, but at a bit rate of 4.5 kbps. This indicates that our proposed APCodec+ is able to synthesize high-quality speech at low bitrate. 

The rest of this paper is organized as follows. In Section \ref{sec:pagestyle}, we give details of our proposed APCodec+. The experimental results and analysis are presented in Section \ref{sec:exp}.
Finally, we give a conclusion in Section \ref{sec:con}.

\vspace{-1mm}
\section{Proposed Method}
\label{sec:pagestyle}
\vspace{-1mm}
% The proposed APCodec+ keep the framework of APCodec \cite{ai2024apcodec} but adopt a novel staged training paradigm. 
% The detailed model structure and training process are discribed as follows. 

\subsection{Model Structure}
\label{ssec:block}
APCodec+ consists of three main components, i.e., encoder, quantizer and decoder. 
The encoder takes amplitude and phase spectra extracted from waveforms using STFT as input and processes them through two parallel sub-encoders. 
These two sub-encoders employ ConvNeXt v2 \cite{woo2023convnext} blocks as the backbone. 
%Initially demonstrating potential in the image domain due to its fewer parameters and strong modeling capabilities, ConvNeXt v2 blocks have been applied in the audio domain \cite{du2023apnet2,ai2024apcodec} as well. 
Finally, features extracted from the amplitude and the phase spectra are downsampled to reduce the frame rate, thereby decreasing the bitrate. 
These features are then concatenated along the feature dimension and passed through dimension-reduction convolutional layers to obtain an intermediate representation with lower dimensionality and frame rate.

The quantizer adopts the RVQ strategy. 
It quantizes the input features through a series of cascaded vector quantizers (VQs). 
The first VQ quantizes the encoded features, while the remaining VQs quantize the quantization errors. 
The final quantized features outputted by the quantizer are obtained by combining the quantization results of all VQs. 
%aiming to minimize the gap between the vectors before and after quantization by constraining the quantization loss term. 
% As suggested in \cite{ai2024apcodec}, the bitrate measured in kbps can be calculated as follows:

% \begin{equation}
%     Bitrate = \frac{1}{1000}\cdot \frac{f_s}{w_s\cdot D}\cdot Q \cdot \log_2 M,
%     \label{eq1}
% \end{equation}
% where $f_s$ is the sampling rate of audio waveform, $w_s$ is the frame shift of the STFT, $D$ is the down/upsampling rate using in the down/upsampling layer, $Q$ is the number of quantizers and $M$ is the size of each codebook in quantizer. A straightforward observation is that using fewer codebooks generally reduces the bitrate, which is also the approach we employed in our experimental section to control the bitrate.

The structure of the decoder mirrors the one of the encoder. 
The quantized features are first dimensionally restored, expanding the spatial range the features can represent. 
They are then injected into two parallel sub-decoders. 
Initially, they undergo upsampling through transposed convolutional layers to enhance the frame rate, followed by a network with ConvNeXt v2 blocks for feature processing. 
Then, these sub-decoders respectively output the decoded amplitude and phase spectra through the linear convolutional layer and phase parallel estimation architecture \cite{ai2023neural}. 
Finally, the decoded audio is reconstructed from the decoded amplitude and phase spectra via iSTFT. 
%Subsequently, they undergo layer normalization and feedforward network processing.

\begin{table*}
	\centering
	\caption{Experiment results of compared intra codecs. \textbf{Bold} indicates the best results at the same bitrate.
}\label{tab1}
	\adjustbox{width=0.9\linewidth}{
 \renewcommand{\arraystretch}{0.9} 

		\begin{tabular}{l c c c c c c}
			\hline
			\hline
			\multirow{1}{*} & { Bitrate (kbps)}& { LSD (dB)$\downarrow$} &{ $\mathrm {AWPD_{IP}}$ (rad)$\downarrow$
}& {$\mathrm {AWPD_{GD}}$ (s)$\downarrow$
} &{$\mathrm {AWPD_{IAF}}$ (rad/s)$\downarrow$
}& { UTMOS$\uparrow$
}\\
			\hline
            { Ground Truth} & -&-&-&-&-&3.579\\
            \hline
            { APCodec} & 6&0.834&1.643&1.375&1.425 & 3.467\\
            { APCodec+} & 6&0.823&1.636&1.371&1.422& \textbf{3.500}\\
            { APCodec*} & 6& \textbf{0.821}& \textbf{1.612}& \textbf{1.365}& \textbf{1.416} & 3.475\\ 
            \hline
            { APCodec} & 4.5&0.844&1.662&1.380&1.429 &3.448\\
            { APCodec+} & 4.5&\textbf{0.841}&\textbf{1.657}&\textbf{1.376}&\textbf{1.426}&\textbf{3.508}\\
            { APCodec*} & 4.5&0.871&1.659&1.380&1.428 &3.256\\ 
            \hline
            { APCodec+} & 3&0.945&1.788&1.402&1.467&3.139\\

			%\hline
			
			\hline
			\hline
	\end{tabular}
}\end{table*}

\subsection{Training Process}
\label{ssec:loss}

Unlike APCodec, the proposed APCodec+ employs a novel staged training paradigm, expanding the traditional joint training mode into a two-stage training paradigm, as shown in Figure \ref{fig2}. 
In the traditional joint training mode, all modules of the codec model are trained jointly. 
One issue with this approach is that, in the early stages of training, the encoder and quantizer are unable to provide high-quality training data for the decoder, making it difficult to train the decoder effectively and limiting the quality and fidelity of the decoded audio to some extent. 
The staged training paradigm introduced in APCodec+ can effectively alleviate this issue. 
The training process of APCodec+ is divided into a joint stage and an individual stage.  

\begin{itemize}[nosep, leftmargin=*]

\item {}{\textbf{Joint Training Stage}}: In the joint training stage, similar to common end-to-end audio codec works, we train the encoder, quantizer, decoder, and discriminator together to achieve joint optimization of all components. 
Through training the whole pipeline, the encoder and quantizer are capable of high-quality feature extraction from the input amplitude and phase spectra, generating quantized feature representations that the decoder can decode into high-quality audio. 
The spectral-level loss, quantization loss and adversarial loss proposed in APCodec \cite{ai2024apcodec} are used in this stage.  
The spectral-level loss includes explicit constraints on amplitude spectrum, phase spectrum, mel spectrogram, and short-time complex spectrum, aiming to ensure the audio fidelity within the spectral domain. 
The quantizer loss aims to minimize the quantization error by defining the mean square error (MSE) between the input and output of each VQ. 
For adversarial training, multi-period discriminator \cite{kong2020hifi} and multi-resolution discriminator \cite{jang2021univnet} are employed to supervise the quality of waveform generation at different periods and resolutions. 
In comparison, AudioDec \cite{wu2023audiodec} omits the adversarial loss during this stage. 
While this enhances training efficiency, it negatively impacts the quality of the quantized features, subsequently affecting the next stage of training.

\item {}{\textbf{Individual Training Stage}}:  In the individual training stage, we individually train the encoder and discriminator from scratch, focusing on enhancing the decoder's modeling capabilities. 
We first freeze the parameters of the encoder and quantizer and produce the quantized features of the training set. 
These quantized features are used as inputs for the decoder, providing training data for both the decoder and the discriminator. 
Then we reinitialize the parameters of the decoder and discriminator randomly, and only train these two modules from scratch. 
During this stage, the quantization loss is discarded.
Compared to the traditional joint training mode, this approach enables the decoder to receive high-quality inputs right from the beginning of training, thereby reducing the training difficulty for both the decoder and the discriminator and accelerating model convergence. 
It resembles the training methods of some adversarial-training-based vocoders, where high-quality speech is synthesized by alternatively updating both the fully initialized parameters of the decoder and discriminator using given natural acoustic features as input. 
Compared with AudioDec \cite{wu2023audiodec}, we can achieve satisfactory performance without the need for additional vocoder assistance.

\end{itemize}

The original intention of the proposed training paradigm is to use sufficiently high-quality quantized features to train the decoder, thereby enhancing its modeling capabilities. 
Therefore, this training paradigm can also be further structured as an iterative process, iteratively executing the aforementioned training procedure. 
Each iteration is expected to generate higher-quality quantized features than the previous one.
However, this approach would increase the model's training cost. 
Some experimental results will be discussed in Section \ref{subsec: Discussion}.

\begin{table*}
	\centering
	\caption{Experiment results of compared inter codecs. \textbf{Bold} indicates the best results.}\label{tab2}
	\adjustbox{width=0.89\linewidth}{
  \renewcommand{\arraystretch}{0.9} 
		\begin{tabular}{l c c c c c c}
			\hline
			\hline
			\multirow{1}{*} & { Bitrate (kbps)}& { LSD (dB)$\downarrow$} &{  $\mathrm {AWPD_{IP}}$ (rad)$\downarrow$
}& { $\mathrm {AWPD_{GD}}$ (s)$\downarrow$
} &{ $\mathrm {AWPD_{IAF}}$ (rad/s)$\downarrow$
}& { UTMOS$\uparrow$
}
\\
			\hline
               { Ground Truth} & -&-&-&-&-&3.579\\

			%\hline
			{ SoundStream} & 12&0.984&1.781&1.415&1.444 &3.023\\
			%\hline			
            { Encodec} & 12&0.933&1.769&1.406&1.439 & 3.318\\
            { HiFi-Codec} & 12&0.858&1.770&1.380&1.432 & \textbf{3.520}\\
            			{ AudioDec} & 12&0.862&1.808&1.426 & 1.456&3.261\\
			{ APCodec+} & 4.5&\textbf{0.841}&\textbf{1.657}&\textbf{1.376}&\textbf{1.426}&3.508\\
			%\hline
			
			\hline
			\hline
	\end{tabular}
}
\end{table*}

\vspace{-1mm}
\section{Experiments}
\label{sec:exp}

\subsection{Experimental Setup}
\vspace{-1mm}
\label{ssec:expset}
\subsubsection{Data and Feature Configuration} 
\vspace{-1mm}
We used Expresso \cite{nguyen2023expresso} for our experiments, which is a multi-speaker English dataset with rich emotion. 
The sampling rate is 48 kHz, and the total duration is approximately 10.87 hours. 
We selected 10,380, 628 and 588 utterances for training, validation and testing, respectively. 
%Due to the small size of the dataset, but rich in style features, we first pre-trained the model on VCTK dataset \cite{yamagishi2019cstr} and then fine-tuned it on Expresso dataset. 
For spectrum extraction, the frame length, frame shift, and the FFT point were set to 320, 40, and 1024 samples respectively. 
%The ratio $D$ of downsampling/upsampling layers in encoder and decoder was set to 8.

%For the quantizer, we set the size of each codebook $M$ to 1024. As calculated by Eq. \ref{eq1}, we adjust the bitrate by changing the number of vector quantizers $Q$, thus the relationship of bitrate (kbps) and $Q$ can be described as: $Bitrate =  1.5Q.$ 

\subsubsection{Implementation}\label{above} 
The APCodec+\footnote{Audio samples can be accessed at \url{ https://redmist328.github.io/APCodecPlus-demo/}.} followed the structural implementation of APCodec \cite{ai2024apcodec}. 
Specifically, the number of ConvNeXt v2 blocks in each backbone network was set to 8.
For each ConvNeXt v2 block, the kernel size and channel size were set to 7 and 512, respectively. 
The dimensionality of the quantizer codebook was set to 32. 
The ratio of downsampling/upsampling layers in the encoder and decoder was set to 8. 
For the quantizer, we set the size of each codebook to 1024.

During training, we randomly cropped the audio clips to 7960 samples and set the batch size to 16.
The model was optimized using the AdamW optimizer \cite{loshchilov2018decoupled} with $\beta_1=0.8$ and $\beta_2 = 0.99$. The learning rate was set initially to $2 \times 10^{-4}$ and scheduled to decay with a factor of 0.999 at every epoch. 
The above configurations were applied to both stages of training.
%Other baseline models were also fine-tuned first on the VCTK dataset before undergoing further fine-tuning on the Expresso dataset.

\subsubsection{Baselines} 
We divided the experiments into two parts, i.e., intra-model evaluation and inter-model evaluation. 
Intra-model evaluation mainly focuses on the improvement of the backbone APCodec \cite{ai2024apcodec} with the proposed staged training paradigm. 
For baseline APCodec, the excellent performing bitrate was 6 kbps as suggested in \cite{ai2024apcodec}. 
Therefore, based on this, we compared the performance of APCodec+ and baseline APCodec at 6 kbps. 
Additionally, to assess how many bitrates this training paradigm saved, we also compared the APCodec+ at 4.5 kbps and 3 kbps with baseline APCodec at 6 kbps. 
Since AudioDec \cite{wu2023audiodec} also proposes a two-stage training mode, it is essential to compare our proposed staged training paradigm with their approach. 
To ensure fairness, we replaced the training framework of APCodec+ with that of AudioDec, denoted as APCodec*, and compared it with APCodec+ at 6 kbps and 4.5 kbps bitrates.
%Finally, we compared our proposed strategy with the training strategy of AudioDec, referred to as APCodec, and conducted experiments at 6kbps and 4.5kbps.

Inter-model evaluation is conducted to compare our APCodec+ with other well-known baseline audio codecs, including SoundStream, Encodec, HiFi-Codec, and AudioDec. 
Their encoded features have a sampling rate that is 1/320 of the waveform's sampling rate. 
They also used a codebook size of 1024.
To demonstrate the compression ability of the proposed APCodec+ and evaluate the bitrate savings, we set the bitrate for these baseline codecs to 12 kbps. 
It should be noted that we adjusted the bitrate of all codec models by changing the number of vector quantizers $Q$.
Under the current configuration, the relationship of the bitrate in kbps and $Q$ can be described as: $Bitrate =  1.5Q$. 

\begin{figure*}[!h]
	\centering
	\includegraphics[width=0.9\linewidth]{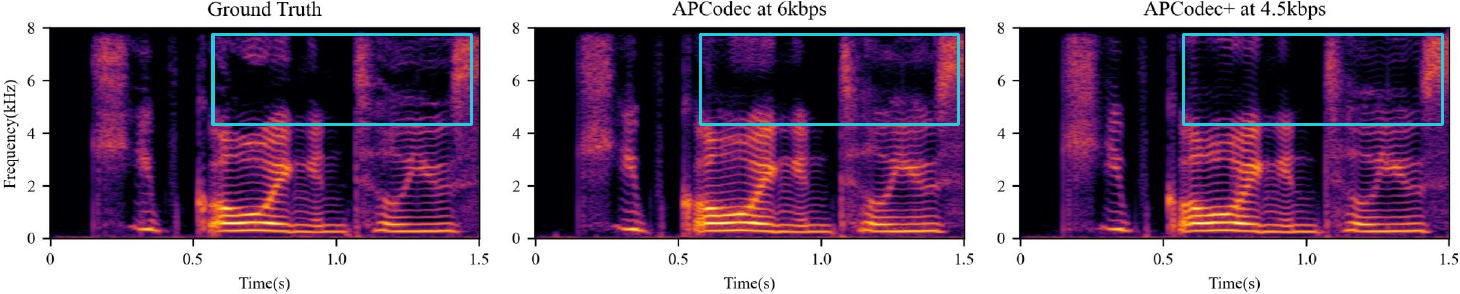}
	\caption{Spectrograms (0$\sim$8 kHz) of ground truth audio and audios decoded by APCodec at 6 kbps and APCodec+ at 4.5 kbps.
 % The selected segment from the sentence (ex04\_sad\_00379.wav) in the test set was used to plot the magnitude spectrum in the frequency range of 0 to 8 kHz.
} \label{fig3}
\end{figure*}
\subsubsection{Evaluation metircs}

We evaluated the above codecs from three aspects, i.e., amplitude spectrum quality, phase spectrum quality, and overall audio quality. 
The log-spectral distance (LSD), a series of anti-wrapping phase distances \cite{ai2024apcodec,lu2024towards} (including instantaneous phase distance $\mathrm {AWPD_{IP}}$, group delay distance $\mathrm {AWPD_{GD}}$, and instantaneous angular frequency distance $\mathrm {AWPD_{IAF}}$), and UTMOS \cite{saeki2022utmos} which is an objective mean opinion score (MOS) evaluation tool with scores ranging from 0 to 5, were used.

% We pay particular attention to indicators of reconstructed speech in both amplitude and phase spectra. Specifically, we utilize the log-spectral distance to measure the distance between the amplitude spectra of real and synthesized waveforms. Additionally, we use previously established metrics for phase evaluation in \cite{ai2024apcodec}, namely instantaneous phase, group delay, and instantaneous angular frequency (referred to as $\mathrm {AWPD_{IP}}$, $\mathrm {AWPD_{GD}}$, and $\mathrm {AWPD_{IAF}}$, respectively). To gauge differences in synthesized speech perception across various systems, we introduce UTMOS \cite{saeki2022utmos} as a subjective mean opinion score evaluation tool. UTMOS is an advanced neural network tool capable of predicting human ratings of speech, with scores ranging from 0 to 5, where higher scores reflect better speech quality. %Model complexity (excluding the discriminators) is also one of the metrics we measure. When used on edge devices, lightweight model parameters are crucial.

\vspace{-1mm}
\subsection{Results and Analysis}

\subsubsection{Intra-Model Evaluation}
\label{ssec:eva}
\vspace{-1mm}
The experimental results of intra-model evaluation are shown in Table \ref{tab1}. 
It is evident that at the optimal bitrate of APCodec (i.e., 6 kbps), both the training strategies proposed in APCodec+ and AudioDec were effective, according to the results of APCodec+ and APCodec*. 
This indicates that the two-stage training mode is beneficial for improving the fidelity of the decoded audio. 
However, the amplitude and phase spectrum quality of APCodec+ was slightly inferior to that of APCodec*. 
Interestingly, as the bitrate decreased to 4.5 kbps, our proposed APCodec+ continued to maintain impressive performance, with the UTMOS score even slightly higher than that at 6 kbps. 
However, all metrics for APCodec* notably decreased, even falling below the performance of the original APCodec at 4.5 kbps. 
This may be attributed to the fact that APCodec* produces unsatisfactory quantized features at low bitrates without adversarial loss, which severely impacts the decoder's learning. 
By comparing APCodec at 6 kbps with APCodec+ at 6 kbps, 4.5 kbps, and 3 kbps, it can be observed that our proposed staged training paradigm can save at least 1.5 kbps in bitrate (equivalent to 1.3$\times$ increase in compression rate) within the framework of APCodec. 
We also compared the spectrograms of decoded audio from APCodec+ at 4.5 kbps and APCodec at 6 kbps in Figure \ref{fig3}. 
As indicated by the blue box, the introduction of the staged training paradigm aids in learning acoustic details (e.g., harmonics). 
This improvement is likely due to the satisfactory quantized features which enhanced the decoder's ability for diverse learning. 
The above experimental results fully confirm that APCodec+ could restore high-fidelity audio and achieve high compression rates.

% From Tab. \ref{tab1}, it is evident that at 6 kbps, both our proposed training strategy and the training strategy of AudioDec resulted in improvements to the original model. Our strategy showed a more noticeable enhancement in the UTMOS perceptual metric, while AudioDec's method exhibited a more significant increase in spectral-level loss. As the bit rate decreased to 4.5 kbps, our method continued to maintain good performance, with the UTMOS score even slightly higher than at 6 kbps. However, all metrics for APCodec* notably decreased, even falling below the performance of the original APCodec at 4.5 kbps. From the magnitude spectrum plotted in Fig. \ref{fig3}, we can observe that our proposed APCodec+ outperforms APCodec in terms of high-frequency harmonic restoration. This demonstrates that our proposed training strategy can achieve high-quality audio generation at lower bitrates, while AudioDec's training strategy does not perform well at lower bit rates. Yet, as the bitrate further decreased to 3 kbps, the effectiveness of our APCodec+ also declined.

\begin{table}
	\centering
	\caption{UTMOS results of APCodec+ at 4.5 kbps when applying iterative training.}\label{tab3}
	\adjustbox{width=0.9\linewidth}{
  \renewcommand{\arraystretch}{1} 
		\begin{tabular}{c c c}
			\hline
			\hline
			No Iteration & One Iteration & Two Iterations\\
   \hline
   \textbf{3.508} &3.492  & 3.480 \\	
			\hline
			\hline
	\end{tabular}}
\end{table}

\vspace{-1mm}
\subsubsection{Inter-Model Evaluation}
\vspace{-1mm}
%\begin{table}
	%\centering
	%\caption{Experiment results of spectrum quality and UTMOS on the test set of Expresso dataset for inter models.}\label{tab2}
	%\adjustbox{width=\linewidth}{
  %\renewcommand{\arraystretch}{1.2} 
	%	\begin{tabular}{l c c c c c c c}
		%	\hline
			%\hline
			%\multirow{2}{*} & { Bitrate}& { LSD} &{  IP
%}& { GD
%} &{ IAF
%}& { UTMOS$\uparrow$
%}& { Model Size
%}\\ &  {  {(kbps)} }& { {(dB)$\downarrow$}}&{ {(rad)$\downarrow$}} &{ { (s)$\downarrow$}}&{ (rad/s)$\downarrow$}&&{ (MB)$\downarrow$}\\
%			\hline
 %              { Ground Truth} & -&-&-&-&-&3.579&-\\

			%\hline
	%		{ SoundStream} & 12&0.984&1.781&1.415&1.444 &3.023&83.2\\
			%\hline			
     %       { Encodec} & 12&0.933&1.769&1.406&1.439 & 3.318&83.2\\
            %{ HiFi-Codec} & 12&0.858&1.770&1.380&1.432 & \textbf{3.520}&243\\
            	%		{ AudioDec} & 12&0.862&1.808&1.426 & 1.456&3.261&108\\
			%{ APCodec+} & 4.5&\textbf{0.841}&\textbf{1.657}&\textbf{1.376}&\textbf{1.426}&3.508&\textbf{66.0}\\
			%\hline
			
			%\hline
		%	\hline
	%\end{tabular}}
%\end{table}

The experimental results of inter-model evaluation are shown in Table \ref{tab2}. 
It can be observed that our proposed APCodec+ at 4.5 kbps outperformed other baselines at 12 kbps in terms of spectral metrics, but it was slightly inferior to HiFi-Codec in terms of perceived overall quality according to UTMOS scores. 
Therefore, our proposed APCodec+ can achieve comparable or better performance at a bitrate of 4.5 kbps compared to some baseline codecs (i.e., SoundStream, Encodec, HiFi-Codec and AudioDec) at 12 kbps, saving 7.5 kbps in bitrate. 
This is attributed to our introduced staged training paradigm, which significantly improved the compression rate while ensuring the fidelity of the decoded audio.

\vspace{-1mm}
\subsubsection{Discussion on Iterative Training}
\label{subsec: Discussion} 
\vspace{-1mm}
As mentioned in Section \ref{ssec:loss}, the proposed staged training paradigm can also be structured as an iterative training mode. 
We iteratively applied the proposed staged training paradigm once and twice by fine-tuning the APCodec+ model at 4.5 kbps. 
The UTMOS results are shown in Table \ref{tab3}. 
We can observe that continuing with iterative training under the proposed staged training paradigm did not further enhance performance but increased the training cost. This indicates that the quality of the quantized features extracted by APCodec+ is sufficient, and there is no need to introduce an additional iterative mode.

\vspace{-1mm}
\section{Conclusion}
\vspace{-1mm}
\label{sec:con}
In this paper, we introduce APCodec+, an improved version of APCodec. 
The highlight of APCodec+ is the introduction of the staged training paradigm. 
The initial joint training stage aims to provide sufficiently high-quality quantized features, while the subsequent individual training stage focuses on enhancing the decoder's modeling capabilities. 
Experimental results confirm that APCodec+ can further improve the compression rate while ensuring the fidelity of the decoded audio. 
Exploring other training methods and introducing APCodec+ into downstream tasks will be our future work.
% We employ a staged training strategy, where the joint training stage involves complete joint training, followed by fixing the parameters of the encoder and quantizer, and randomly initializing the parameters of the decoder and discriminator. The individual training stage focuses solely on training the parameters of the decoder and discriminator. These experiments demonstrate that APCodec+ can maintain high-quality speech generation at low bit rates and significantly save bitrates compared to other baseline models.

\bibliographystyle{IEEEtran}

\bibliography{mybib}

% \begin{thebibliography}{9}
% \bibitem[1]{Davis80-COP}
%   S.\ B.\ Davis and P.\ Mermelstein,
%   ``Comparison of parametric representation for monosyllabic word recognition in continuously spoken sentences,''
%   \textit{IEEE Transactions on Acoustics, Speech and Signal Processing}, vol.~28, no.~4, pp.~357--366, 1980.
% \bibitem[2]{Rabiner89-ATO}
%   L.\ R.\ Rabiner,
%   ``A tutorial on hidden Markov models and selected applications in speech recognition,''
%   \textit{Proceedings of the IEEE}, vol.~77, no.~2, pp.~257-286, 1989.
% \bibitem[3]{Hastie09-TEO}
%   T.\ Hastie, R.\ Tibshirani, and J.\ Friedman,
%   \textit{The Elements of Statistical Learning -- Data Mining, Inference, and Prediction}.
%   New York: Springer, 2009.
% \bibitem[4]{YourName17-XXX}
%   F.\ Lastname1, F.\ Lastname2, and F.\ Lastname3,
%   ``Title of your INTERSPEECH 2022 publication,''
%   in \textit{Interspeech 2022 -- 23\textsuperscript{rd} Annual Conference of the International Speech Communication Association, September 18-22, Incheon, Korea, Proceedings, Proceedings}, 2022, pp.~100--104.
% \end{thebibliography}

\end{document}